\documentclass[reprint,amsmath,amssymb,aps,biblatex,prx]{revtex4-2}

\usepackage{graphicx}% Include figure files
\usepackage{dcolumn}% Align table columns on decimal point
\usepackage{bm}% bold math
\usepackage{lineno}
\usepackage{braket}
\usepackage{svg}
\usepackage{siunitx}
\usepackage{bbm}
\usepackage{multirow}

\begin{document}

%%%%%%%%%%%%%%%%%%%%% numbers for results %%%%%%%%%%%%%%%%%%%%%

\newcommand\FRBthreelevel{0.9951}
\newcommand\deltaFRBthreelevel{0.0002}
\newcommand\FDthreelevel{0.9971}
\newcommand\deltaFDthreelevel{0.0001}

\newcommand\FRBfivelevel{0.9886}
\newcommand\deltaFRBfivelevel{0.0002}
\newcommand\FDfivelevel{0.9931}
\newcommand\deltaFDfivelevel{0.0001}

\newcommand\FRBeightlevel{0.9825}
\newcommand\deltaFRBeightlevel{0.0006}
\newcommand\FDeightlevel{0.9895}
\newcommand\deltaFDeightlevel{0.0003}

%%%%%%%%%%%%%%%%%%%%%%%%%%%%%%%%%%%%%%%%%%%%%%%%%%%%%%%%%%%%%%%

% for writing Clebsch-Gordon coefficients
\newcommand{\cg}[6]{
\left\langle 
\begin{array}{@{}cc|c@{}}
    #1 & #3 & #5 \\
    #2 & #4 & #6
\end{array}
\right\rangle
}
\title{Multi-frequency control and measurement of a spin-7/2 system encoded in a transmon qudit}

\author{Elizabeth Champion}
    \thanks{These authors contributed equally to this work.}
    %\email{elizabeth.champion@rochester.edu}
\author{Zihao Wang}
    \thanks{These authors contributed equally to this work.}
\author{Rayleigh Parker}
\author{Machiel Blok}
    \email{machielblok@rochester.edu}
\affiliation{
 Department of Physics and Astronomy, University of Rochester, Rochester, NY 14627
}

\date{\today}

\begin{abstract}
\noindent Qudits hold great promise for efficient quantum computation and the simulation of high-dimensional quantum systems \cite{wang_qudits_2020}.
Utilizing a local Hilbert space of dimension $d > 2$ is known to speed up certain quantum algorithms relative to their qubit counterparts given efficient local qudit control and measurement.
However, the direct realization of high-dimensional rotations and projectors has proved challenging, with most experiments relying on decompositions of $SU(d)$ operations into series of rotations between two-level subspaces of adjacent states and projective readout of a small number of states \cite{dolde_high-fidelity_2014,asaad_coherent_2020,ringbauer_universal_2022,chi_programmable_2022,bianchetti_control_2010}. Here we employ simultaneous multi-frequency drives to generate rotations and projections in an effective spin-$7/2$ system by mapping it onto the energy eigenstates of a superconducting circuit. 
We implement single-shot readout of the 8 states using a multi-tone dispersive readout ($F_{\rm assignment} = 88.3 \%$) and exploit the strong nonlinearity in a high $E_J / E_C$ transmon to simultaneously address each transition and realize a spin displacement operator. 
By combining the displacement operator with a virtual SNAP gate, we realize arbitrary single-qudit unitary operations in $\mathcal{O}(d)$ physical pulses and extract spin displacement gate fidelities ranging from $0.997$  to $0.989$ for virtual spins of size $j = 1$ to $j = 7 / 2$. 
These native qudit operations could be combined with entangling operations to explore qudit-based error correction \cite{gross_codes_2021, campbell_enhanced_2014} or simulations of lattice gauge theories with qudits \cite{bauer_quantum_2023}. Our multi-frequency approach to qudit control and measurement can be readily extended to other physical platforms that realize a multi-level system coupled to a cavity and can become a building block for efficient qudit-based quantum computation and simulation. 
\end{abstract}

\maketitle
\section{Introduction}

Quantum computing relies on the ability to encode, control and measure quantum information with high precision in a large state space. The canonical approach uses qubits encoded in coupled (effective) spin-1/2 systems that can be controlled with high-fidelity owing to their strong nonlinearity \cite{evered_high-fidelity_2023, chen_benchmarking_2023,acharya_suppressing_2023,xue_quantum_2022}. In contrast, continuous variable encodings use the infinite dimensional state space of a linear bosonic mode, enabling, e.g., resource efficient quantum error correction \cite{ni_beating_2023, sivak_real-time_2023}. Qudits can combine the advantages of both approaches by encoding information in (effective) large spins: finite $d$-level systems ($d>2$) with strong nonlinearity. Qudit algorithms are predicted to outperform their qubit counterparts at quantum information tasks including magic state distillation for quantum error correction \cite{campbell_magic-state_2012} and the synthesis of many-body gates \cite{gokhale_asymptotic_2019}. Furthermore, qudits are promising candidates for the simulation of quantum fields \cite{ciavarella_trailhead_2021, gonzalez-cuadra_hardware_2022,gustafson_prospects_2021} or large-angular momentum spins \cite{catarina_hubbard_2022, gong_kaleidoscope_2016} owing to their natural correspondence with high-dimensional quantum systems.  

Experimental qudit processors have recently been demonstrated using high-dimensional nuclear spins \cite{dolde_high-fidelity_2014, asaad_coherent_2020}, hyperfine states of trapped ions \cite{ringbauer_universal_2022}, photonic circuits \cite{chi_programmable_2022}, and superconducting circuits \cite{blok_quantum_2021}. So far readout has been limited to sequential qubit-like readout, or limited to up to four states simultaneously \cite{peterer_coherence_2015, xian_software_2020, chen_threshold_2023, nguyen_empowering_2023, kehrer_ibm_2024, cao_emulating_2024}. Universal qudit control in these processors is typically realized by sequentially applying Givens rotations between adjacent energy states requiring $\mathcal{O}(d^2)$ qubit-like rotations to execute arbitrary single-qudit gates \cite{liu_performing_2023, morvan_qutrit_2021}. However, this approach leaves most energy states idling during gates and can lead to complex error propagation. 
Thus far, experiments that explored simultaneous drives on multiple transitions, either with optimal control pulses \cite{seifert_ququart_2023}, or with known decompositions \cite{yurtalan_hadamard_2020, neeley_emulation_2009} were limited to three- or four-level systems, without a clear path towards high-fidelity qudit rotations in larger dimensions.
A platform capable of implementing native multi-level rotations and readout would unlock the power of qudits by making efficient use of the larger Hilbert space and allowing a direct mapping between the platform native operations and qudit algorithms or simulations.

In this work we realize spin displacement operations in a superconducting transmon qudit with up to $d = 8$ levels.
This is facilitated by a modified transmon design with a deep cosine potential which encodes qudit states with coherence times close to the lifetime limit, enabling high-fidelity control.
Extending qubit dispersive cavity readout to a multi-tone measurement we realize simultaneous single-shot readout of the eight energy states with average assignment fidelity of 88 percent.
Unlike their bosonic counterpart, spin systems are inherently finite-dimensional, and have a spherical rather than planar phase space.
The family of spin displacement unitaries likewise consists of rotations on this phase space.
We leverage the strong nonlinearity of the transmon to realize spin displacement operators by simultaneously driving the individual transitions.
Utilizing this family of displacement operations, we perform direct Wigner tomography of the qudit state and verify the Wigner function's expected behavior.
Furthermore, we prepare spin cat states by combining the displacement operator with a virtual selective number-dependent arbitrary phase (SNAP) gate, which is implemented with close to unity fidelity by updating the respective drive pulse phases. 
Finally, we demonstrate a gate set consisting of a single displacement pulse and this virtual SNAP gate to do arbitrary qudit rotations in $\mathcal{O}(d)$ physical pulses, reaching native gate fidelities of $F_{\hat{D}} = \{\num{\FDthreelevel \pm \deltaFDthreelevel}, \num{\FDfivelevel \pm \deltaFDfivelevel}, \num{\FDeightlevel \pm \deltaFDeightlevel} \}$ for $d = 3$, $d = 5$, and $d = 8$ respectively.

\begin{figure}
    \includegraphics[width=1\columnwidth]{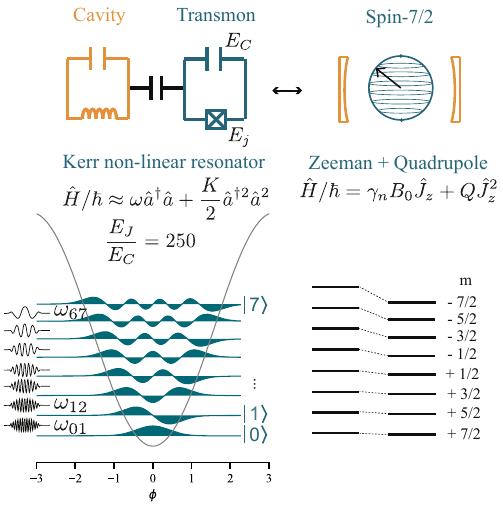}
    \caption{Encoding large spins in high-$E_J / E_C$ transmon qudits.
    Mapping a large spin onto a transmon. We dispersively couple a transmon with $E_J/ E_C \approx 250$ to a readout resonator. The resonator enables projective readout of the qudit state, while the large $E_J / E_C$ increases the number of states confined in the cosine potential and suppresses charge noise in the highly excited states. We map the Hilbert space of a large spin onto the transmon qudit such that the spin's $\hat{J}_z$ eigenstates correspond to the transmon eigenstates. Under this mapping our system resembles a large spin dispersively coupled to a cavity, with a small quadrupole term corresponding to the lowest-order (Kerr) nonlinearity of the transmon.
    }
    \label{fig:overview}
\end{figure}

\section{Results}

%
% Transmon qudits
%

The fixed-frequency superconducting transmon oscillator is often used as a physical qubit for quantum information processing \cite{koch_transmon_2007}.
It can be represented as a lumped-element circuit consisting of a capacitor with characteristic charging energy $E_C$ and a nonlinear inductance in the form of a Josephson junction of characteristic energy $E_J$.
The resulting circuit behaves as an anharmonic oscillator, with a Kerr nonlinearity as the lowest-order correction to the harmonic potential.
The two lowest-lying energy eigenstates are typically used to encode a qubit in an effective spin-$1/2$ system.
Typical ratios of $E_J / E_C \approx 50$ suppress charge noise-induced dephasing within this qubit subspace while producing an appreciable anharmonicity of around $\qty{300}{\mega\hertz}$ \cite{koch_transmon_2007, blais_cqed_2021}.
The qubit is controlled using a resonant microwave-frequency drive to address the $\ket{0} \leftrightarrow \ket{1}$ transition.
Readout is enabled by dispersively coupling the qubit to an $LC$ oscillator realized as a coplanar waveguide resonator.
The lumped-element representation of this circuit is shown in Figure \ref{fig:overview}.

While the two lowest-energy eigenstates are well-suited for qubit computation, one can include the higher excited states to realize a qudit \cite{blok_quantum_2021, morvan_qutrit_2021, liu_performing_2023, nguyen_empowering_2023}.
The anharmonicity due to the Josephson junction ensures that each of the excited state transitions is spectrally resolved.
Typical $E_J / E_C$ ratios permit the use of a few excited states beyond the qubit subspace; however, the number of states confined within the Josephson potential is relatively small, and the higher states are subject to appreciable charge noise\cite{koch_transmon_2007}.
We address both shortcomings with a modified device design having $E_J / E_C \approx 250$.
The cosine potential and phase-basis energy eigenstate wavefunctions are shown in the lower panel of Figure \ref{fig:overview}.
In the following experiments we use up to seven excited states for qudit computation in $d = 8$ dimensions.
The Hilbert space of an 8-level qudit is equivalent to an effective spin-$7/2$ system, and the spectral separation of our eigenstates can be seen as analogous to a high-dimensional spin with a quadrupole term in its Hamiltonian.
In this picture the harmonic part of the transmon potential, $\hat{a}^\dag \hat{a}$, plays the role of a Zeeman Hamiltonian $\gamma_n B_0 \hat{J}_z$ due to an applied magnetic field, while the Kerr term $\hat{a}^{\dag2} \hat{a}^2$ corresponds to the spin quadrupole term $Q \hat{J}_z^2$.
In a transmon the anharmonicity across all transitions is approximately given by $K \approx E_C$, where in our device we estimate $E_C = \qty{108}{\mega\hertz}$.
%In a transmon the anharmonicity across all transitions, corresponding to this Kerr nonlinearity, is approximately given by the charging energy $E_C$, which we estimate to be $\qty{108}{\mega\hertz}$.
Including the readout resonator one can imagine our system as a large spin dispersively coupled to a cavity, where the $\hat{J}_z$ eigenstates are mapped onto the transmon eigenstates (Figure \ref{fig:overview}, right).

\begin{figure*}
    \includegraphics{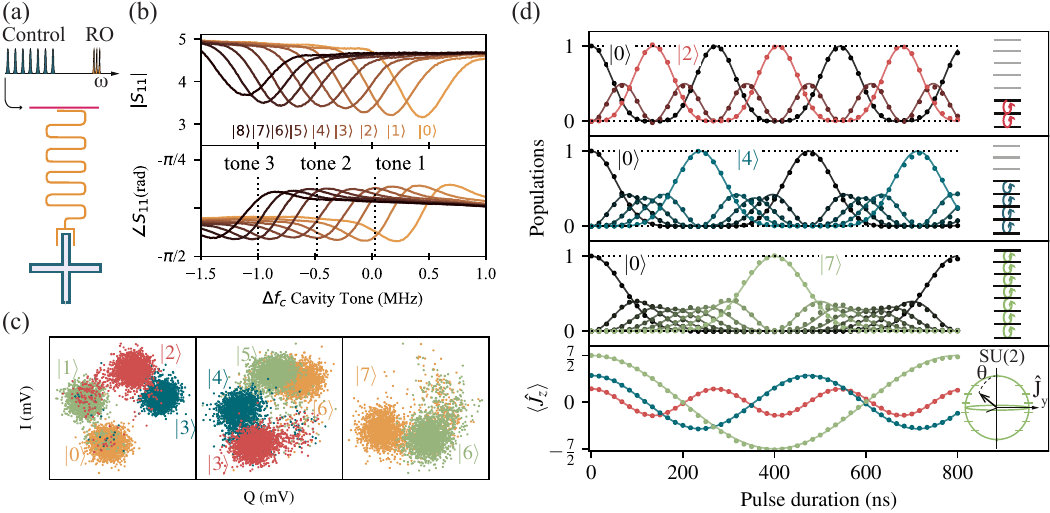}
    \caption{Implementing control and measurement with multi-tone drives.
    (a) Illustration of the high-$E_J/E_C$ fixed-frequency transmon and dispersively coupled readout resonator. Transmon drives are applied through the readout line.
    (b) The readout resonator spectrum for each transmon eigenstate. We perform projective readout of the qudit state by multiplexing three readout tones at different frequencies such that each state can be distinguished from the others by at least one tone.
    (c) Qudit readout in the $IQ$ plane of each readout tone.
    (d) Implementing the spin displacement operator by simultaneously driving multiple transitions. The pulse amplitudes are scaled such that the largest element of the displacement generator matrix is held roughly fixed, producing slower oscillations as the number of dimensions is increased.
    The solid lines show single-parameter fits to the experimental data accounting for an overall scale factor on the pulse amplitude, of order unity.
    We also compute the expectation value of the $\hat{J}_z$ angular momentum operator for each case (bottom).
    The displacement drive induces sinusoidal oscillations in the angular momentum, realizing a complete flip of the angular momentum when the displacement angle is $\theta = \pi$.
    }
    \label{fig:time_sweep}
\end{figure*}

As in the case of a qubit, the dispersively coupled resonator facilitates readout of the transmon state.
In the present case, however, we have optimized the device design to permit single-shot readout of multiple levels.
We show the dispersive shift of the readout resonator spectrum in Figure \ref{fig:time_sweep}b for transmon states $\ket{0}$ through $\ket{8}$.
A single-frequency readout tone is able to distinguish several of these states, limited by the linewidth of the resonator relative to the dispersive shift.
We realize single-shot readout of the entire qudit by multiplexing three readout tones with frequencies chosen such that together they can distinguish each state from the others \cite{chen_threshold_2023}.
We demodulate the signal reflected from the resonator at each of these three frequencies and integrate each to produce three pairs of IQ values.
We consider these as a single point in a 6-dimensional space, which is classified to determine the readout result.
Altogether this results in an assignment fidelity of $(88.3 \pm 0.2) \%$; see the Supplemental Material \cite{supp} for details.

%This projective qudit readout allows us to visualize the spectrum of the transmon itself, shown in Figure \ref{fig:overview}c.
%We prepare an equal superposition of the even-parity eigenstates using a series of pulses between adjacent eigenstates, and following this we sweep over the frequency and duration of an applied drive tone.
%We then perform readout, and from many repetitions of the experiment we reconstruct the expectation value of the eigenstate parity operator.
%The result is a series of chevron-like patterns showing Rabi oscillations when the drive tone is resonant with one of the qudit transitions, illustrating the coherent control enabled by the spectral separation of the transitions.

The spectral separation of the qudit transitions allows us to selectively apply rotations between adjacent states $\ket{n-1}$ and $\ket{n}$, given that the drive strength $\Omega$ is considerably less than the anharmonicity, which in our device is $E_C = \qty{108}{\mega\hertz}$.
%which in our device is $\alpha = \qty{114}{\mega\hertz}$.
In combination with the qudit readout this enables coherence time measurements of each transition.
%To measure the $T_1$ of a given transition $\ket{n} \rightarrow \ket{n-1}$ we apply a series of $\pi$ pulses, beginning in the ground state, in order to prepare the state $\ket{n}$.
%Allowing the state to decay for a variable time before performing readout, we measure the per-transition decay rate and find $T_1^{01} = \qty{46}{\micro\second}$ down to $T_1^{67} = \qty{13}{\micro\second}$.
We measure the energy relaxation rate $T_1^{(n-1)(n)}$ for each given transition $\ket{n} \rightarrow \ket{n-1}$  and find $T_1^{01} = \qty{46}{\micro\second}$ down to $T_1^{67} = \qty{13}{\micro\second}$, where the increased relaxation for the higher transitions is largely explained by the scaling of the matrix elements of the lowering operator, analogous to bosonic enhancement for harmonic modes.
We perform similar measurements of $T_2$ and $T_2^*$ using $\pi/2$ pulses to perform echo and Ramsey experiments within each subspace and find dephasing times ($T_2$) close to the lifetime limit; see the Supplemental Material \cite{supp} for the full set of coherence times.

%
% Explain spin displacements
%

With our coherent, transition-selective control in mind, we now consider the implementation of a spin qudit displacement operator, beginning with a brief overview of high-dimensional spins and their displacements.
The Hilbert space of a spin-$j$ system is spanned by a basis of $d = 2j + 1$ states, usually labeled $\ket{j,m}$ for $-j \leq m \leq j$.
These can be taken to be the eigenstates of the angular momentum operator $\hat{J}_z$.
The dynamical symmetry group of a spin of any dimension is $SU(2)$ with corresponding Lie algebra $\mathfrak{s}\mathfrak{u}(2)$.
The representation of this Lie algebra in a spin-$j$ system consists of $d \times d$ Hermitian matrices, spanned by the basis $\{\hat{J}_x, \hat{J}_y, \hat{J}_z \}$.
The displacement operator of a spin-$j$ system is thus an element of the $SU(2)$ representation generated by this basis:
\begin{equation}
    \hat{D}(\theta, \phi) = e^{-i \phi \hat{J}_z} e^{-i \theta \hat{J}_y} = e^{\alpha \hat{J}_+ - \alpha^* \hat{J}_-},
\end{equation}
where the spin ladder operators are $\hat{J}_\pm \equiv \hat{J}_x \pm i \hat{J}_y$ and $\alpha \equiv -\frac {\theta} {2} e^{-i \phi}$ \cite{perelomov_generalized_1977, davis_wigner_negativity_2021}.
The coherent states of a high-dimensional spin are those generated by applying the displacement operator to a fixed state in the Hilbert space, usually taken to be the ground state: $\ket{\theta,\phi} = \hat{D}(\theta, \phi) \ket{j,j}$.
As in the case of a harmonic oscillator, applying a displacement operator to a coherent state yields another coherent state; furthermore, the coherent states form an overcomplete basis for the Hilbert space.
The phase space representation of a spin is closely related to these coherent states.
The phase space itself is the surface of a sphere, and the displacement angles $\theta$ and $\phi$ specify polar and azimuthal angles on the phase space sphere, respectively.

%
% Mapping between the Hilbert spaces
%

We translate this picture to our transmon qudits by mapping the spin state $\ket{j,m}$ onto the transmon oscillator state $\ket{n} = \ket{j-m}$.
The ground state of the spin system, assuming a Zeeman Hamiltonian proportional to $-\hat{J}_z$, is $\ket{j,j}$, which under this mapping corresponds to the transmon's ground state, $\ket{0}$.
In this notation the action of a spin ladder operator is to increment the number of excitations in the transmon: $\hat{J}_+ \ket{n} = \sqrt{n (d - n)} \ket{n-1}$ and $\hat{J}_- \ket{n} = \sqrt{(n + 1)(d - n - 1)} \ket{n+1}$.
We note that in the limit of $n \ll d$ these ladder operators approach the bosonic operators $\hat{a}$ and $\hat{a}^\dag$ respectively, up to a multiplicative constant.

%
% Driving displacements
%

We implement high-dimensional spin displacements in transmons by exploiting the nonlinearity of the system to simultaneously drive multiple transitions\cite{neeley_emulation_2009}.
We first note that, under the rotating wave approximation (RWA), a single resonant drive at the $\ket{n-1} \leftrightarrow \ket{n}$ transition frequency produces an effective $\hat{\sigma}_x$-like Hamiltonian within that subspace, thus inducing Rabi oscillations between those levels.
The validity of the RWA is contingent on the applied drive strength being appreciably smaller than the anharmonicity of the qudit ($\Omega \ll K)$.
We calibrate the drive strength necessary to reach a given Rabi rate separately for each transition in our Hilbert space.
Because the spin ladder operators $\hat{J}_\pm$ connect only adjacent states, we observe that we can drive displacement operations by simultaneously driving every transition.
The drive strengths are scaled such that, if the drive on $\ket{n-1} \leftrightarrow \ket{n}$ were applied on its own, it would yield Rabi oscillations of Rabi frequency $\Omega(t) \sqrt{n (d - n)}$ for some global time-dependent envelope $\Omega(t)$.
The total displacement induced by these drives after a time $T$ is $\hat{D}(\theta,\phi)$ where $\theta = \int_0^T \Omega(t) dt$ and $\phi$ is the phase applied to each drive.
We use a flat pulse with cosine ramps at the beginning and end, each having a duration of $1/4$ of the total pulse length.
%We note that similar experiments were carried out with fewer qudit states in Ref. \cite{neeley_emulation_2009} using a superconducting phase qudit.

Applying the displacement drive to the ground state produces spin coherent states, shown in Figure \ref{fig:time_sweep}d for several different Hilbert space dimensions.
We note that the dimensionality of the emulated spin is artificially imposed by restricting our drives to the appropriate subset of transitions.
After applying a displacement pulse of a given duration we perform qudit readout and plot the populations.
We scale the drive amplitudes between dimensions such that the largest matrix element of the displacement generator is kept roughly fixed, producing slower oscillations as the number of dimensions is increased.
Because we have chosen to map $\hat{J}_z$ eigenstates onto the transmon eigenstates, we can directly compute the expectation value $\langle \hat{J}_z \rangle$ from the populations, shown in the bottom panel.
When the area of the displacement pulse is $\theta = \pi$ the population is transferred entirely to the highest excited state, $\ket{d-1}$, corresponding to a full flip of the effective angular momentum.
These results illustrate that we can drive high-fidelity $SU(2)$ rotations of emulated spins of varying dimension, limited only by the number of qudit states available in our system.

%
% Wigner functions
%

\begin{figure*}
    \includegraphics{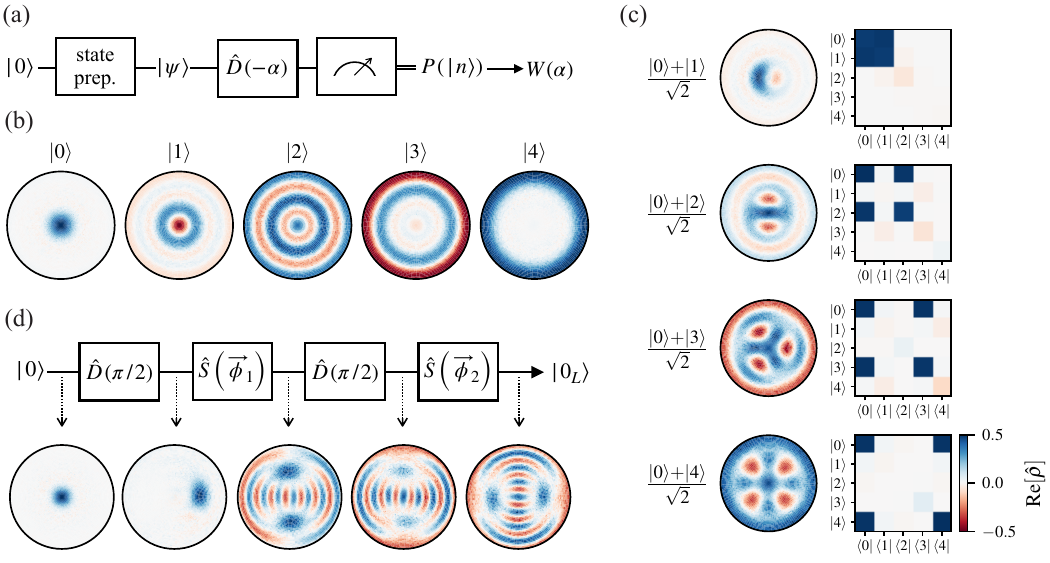}
    \caption{Representing transmon qudit states in phase space.
    We plot a polar projection of the $SU(2)$ phase space sphere onto a disk such that the radial coordinate is $\theta$ and the angular coordinate is $\phi$;
    in this projection the north pole of the $SU(2)$ sphere lies at the center of the disk while the south pole is spread across its boundary.
    (a) Pulse sequence for Wigner tomography. Beginning in the ground state, the desired state is constructed using a series of $\pi$ pulses for (b) and a $\pi/2$ pulse followed by $\pi$ pulses for (c). Following this, a displacement drive is applied and the qudit state is measured. Populations estimated from many repetitions of this sequence are used to construct the Wigner function.
    (b) Measured Wigner functions for energy eigenstates in $d = 5$ dimensions.
    (c) Measured Wigner functions for equal superpositions of the ground state and each excited state. To the right we show the real parts of the density matrices recovered from these Wigner functions; we attribute discrepancies between these density matrices and the ideal density matrices primarily to imperfections in the calibration of the Wigner scan's displacement pulse, which we expect can be further improved in future work.
    (d) SNAP-displacement sequence for encoding a cat state in $d = 8$.
    The sequence maps $\ket{0}$ onto the cat state $(\ket{\frac {\pi} {2}, 0} + \ket{-\frac {\pi} {2}, 0}) / \sqrt{2}$, which we use as the logical qubit state $\ket{0_L}$.
    We omit the initial SNAP gate layer since in this case it acts on $\ket{0}$ and thus has no physical effect.
    } \label{fig:wigner}
\end{figure*}

One application of this $SU(2)$ displacement operation is the direct measurement of the spin qudit Wigner function.
Given a density matrix $\hat{\rho}$, the Wigner function at a phase space coordinate $\alpha$ is given by $W(\alpha) = \text{Tr}[\hat{\rho} \hat{\Delta}(\alpha)]$, where $\hat{\Delta}(\alpha)$ is an operator-valued function over phase space known as the kernel.
The appropriate kernel for a given system is not unique, needing only to satisfy the Stratonovich-Weyl postulates \cite{stratonovich_distributions_1957, heiss_discrete_2000, davis_wigner_negativity_2021}.
In the case of a spin it can be shown that one such kernel has the form of a displaced parity-like operator, in analogy with the harmonic oscillator: $\hat{\Delta}(\alpha) = 2 \hat{D}(\alpha) \hat{\Pi} \hat{D}(-\alpha)$ where $\hat{\Pi}$ is diagonal in the $\hat{J}_z$ basis \cite{supp}.
We measure the Wigner function at a point $\alpha$ by applying a displacement $\hat{D}(-\alpha)$ and performing qudit readout to extract the expectation value of $\hat{\Pi}$.
We demonstrate this in Figure \ref{fig:wigner}b and \ref{fig:wigner}c for energy eigenstates and superposition states, respectively. 
As in the case of a harmonic oscillator, energy eigenstates are completely delocalized in phase, producing rotationally symmetric phase space distributions.
The ground state and highest excited state are themselves examples of spin coherent states and therefore have nearly entirely positive Wigner functions, while the other energy eigenstates have large Wigner-negative regions indicative of nonclassicality.

%
% RB figure
%

\begin{figure*}
    \includegraphics{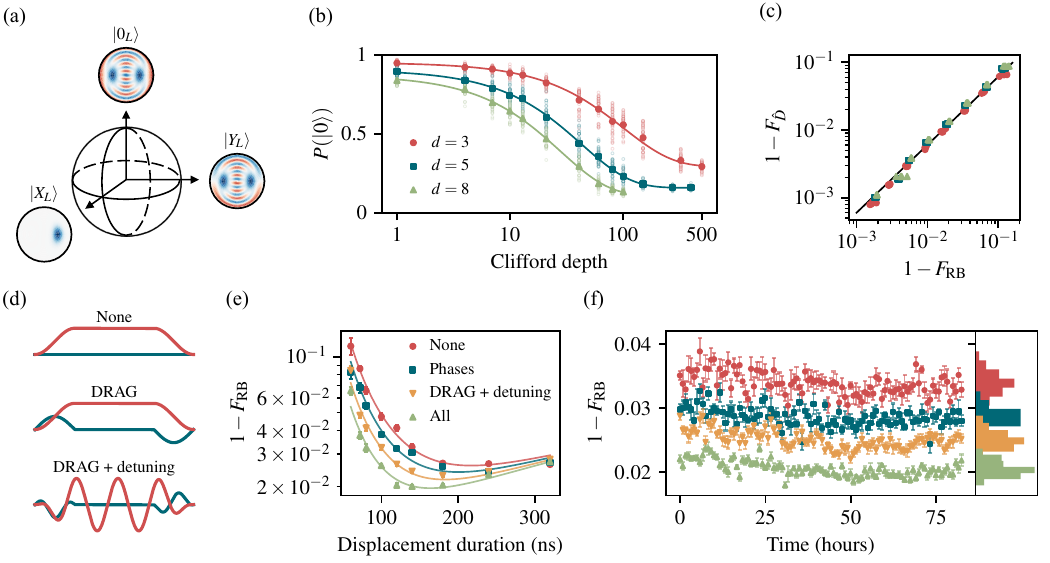}
    \caption{Randomized benchmarking of a logical qubit controlled with SNAP-displacement decompositions.
    (a) The logical Bloch sphere with theoretical Wigner functions in $d = 8$.
    (b) Randomized benchmarking in $d = 3$, $d = 5$, and $d = 8$. %The displacement pulse lengths are chosen to maximize the fidelity, and the pulses include DRAG, detuning, and phase corrections determined from numerical simulations with an optimal control code. 
    We extract average Clifford gate fidelities of $F_{\rm RB} = \{\num{\FRBthreelevel \pm \deltaFRBthreelevel}, \num{\FRBfivelevel \pm \deltaFRBfivelevel}, \num{\FRBeightlevel \pm \deltaFRBeightlevel} \}$ respectively, yielding estimated displacement fidelities of $F_{\hat{D}} = \{\num{\FDthreelevel \pm \deltaFDthreelevel}, \num{\FDfivelevel \pm \deltaFDfivelevel}, \num{\FDeightlevel \pm \deltaFDeightlevel} \}$.
    (c) Validating displacement fidelities extracted from randomized benchmarking. We numerically simulate our randomized benchmarking procedure for displacement operations with varying (known) infidelities such that we can relate the underlying displacement fidelities to the Clifford gate fidelities extracted from exponential fits. We find that, regardless of the number of qudit dimensions being used, the results are well-modeled by the relation $F_{\hat{D}} = (F_{\rm RB})^{1/N}$ with $N = 5/3$ being the average number of displacement pulses per Clifford operation (black line).
    (d) The cosine-square pulse shape used for the displacement pulses. The in-phase component of the drive is shown in red, and the out-of-phase component in blue. DRAG-like corrections are calculated numerically, producing some amplitude in the out-of-phase part; detunings, meanwhile, can be viewed as oscillations in the envelopes. These corrections are applied per-transition, i.e., each transition's drive has independent DRAG weights and detunings. Note that we have exaggerated the DRAG weight and detuning to better illustrate their effects on the pulse.
    (e) Randomized benchmarking fidelities in $d = 8$ as a function of displacement pulse duration for each level of correction. The fit model assumes that the total infidelity is the sum of coherent and incoherent parts, with the coherent error decreasing with the square of the pulse duration, reflecting phase errors and leakage due to off-resonant driving, and the incoherent error increasing linearly with duration, due to e.g. energy relaxation. Our model further assumes that all correction settings are subject to the same rate of incoherent error.
    (f) Repeated randomized benchmarking in $d = 8$. We demonstrate the stability of our displacement pulse implementation as well as the significance of the pulse corrections by repeating the same randomized benchmarking experiment many times over the course of 82 hours, without recalibrating, showing the overall distribution of Clifford errors in the right panel.
    } \label{fig:rb}
\end{figure*}

%
% Control in SU(d)
%

In addition to enabling Wigner tomography, we now propose spin displacements as primitive gates for efficient qudit computation.
Qudit gates are typically realized by decomposing the desired unitary into a series of Givens rotations within two-level subspaces of adjacent states $\{ \ket{n-1}, \ket{n} \}$.
Such decompositions require, in general, $d(d-1)/2$ such rotations \cite{liu_performing_2023, nguyen_empowering_2023, deGuise_factorization_2018}, which can become costly for high-dimensional qudits.
Inspired by the SNAP-displacement control scheme for bosonic systems \cite{heeres_snap_2015, krastanov_snap_2015, kudra_negative_2022, ma_control_2021}, we find that interleaving $N$ spin displacement operations with $N + 1$ arbitrary phase rotations provides universal control over the qudit Hilbert space, given sufficiently large $N$ \cite{supp}:
\begin{equation}
    \hat{U}  = \hat{S}(\vec{\phi}^{(N)}) \hat{D}(\theta^{(N)}) \ldots \hat{S}(\vec{\phi}^{(1)}) \hat{D}(\theta^{(1)}) \hat{S}(\vec{\phi}^{(0)}).
\end{equation}.
The required circuit depth for this scheme, $N$, is advantageous as compared to Givens rotations: we find numerically that only $\mathcal{O}(d)$ displacement layers are required to realize arbitrary operations in $SU(d)$.
Moreover, unlike bosonic systems, we implement the SNAP layers simply by updating subsequent drive phases, producing virtual rather than physical phase rotations\cite{mckay_efficient_2017,morvan_qutrit_2021}.
Thus any single-qudit unitary can in principle be realized by $\mathcal{O}(d)$ physical pulses.
Because arbitrary $SU(2)$ rotations can be realized using two $\pi/2$ rotations alternating with $\hat{J}_z$ rotations (which are themselves examples of SNAP gates), we can optionally replace each general $\hat{D}(\theta)$ displacement with two $\hat{D}(\pi/2)$ displacements with a virtual $\hat{J}_z$ rotation between them, at the moderate cost of doubling the number of SNAP-displacement layers.
%impose the restriction that $\theta^{(i)} = \pi/2$ for all $i$, at the moderate cost of doubling the number of SNAP-displacement layers.
We therefore focus our efforts on calibrating high-fidelity $\pi/2$ displacements and decompose arbitrary unitaries into sequences of this single pulse interleaved with virtual phase rotations.
To implement a given target unitary we then find the decomposition parameters $\vec{\phi}^{(i)}$ using numerical optimization. 

We demonstrate the SNAP-displacement gate set in Figure \ref{fig:wigner}d by preparing a spin cat state, i.e., an equal superposition of opposite-phase spin coherent states, in $d = 8$ dimensions.
In particular we choose coherent states lying on the equator of the phase space sphere, ideally preparing the state $( \ket{\frac {\pi} {2}, 0} + \ket{-\frac {\pi} {2}, 0} ) / \sqrt{2}$.
We find in this case that the cat state can be prepared with only two physical displacement pulses. %, and we obtain a state fidelity of $F_{\rm cat} \approx [{\rm number}] \%$.
We note that much like the harmonic oscillator cat states, our spin cat state only contains even-parity eigenstates; the orthogonal cat state $( \ket{\frac {\pi} {2}, 0} - \ket{-\frac {\pi} {2}, 0} ) / \sqrt{2}$ likewise only contains odd-parity eigenstates.
It is important to remember, however, that our cat state is a superposition of spin coherent states rather than harmonic oscillator coherent states, differing both in the amplitudes of the eigenstates as well as the overall dimensionality of the system \cite{supp}.

While the SNAP-displacement decomposition is universal over $SU(d)$, benchmarking its performance over the full Hilbert space presents a challenge.
In principle one could perform qudit randomized benchmarking \cite{morvan_qutrit_2021} where each qudit Clifford gate is decomposed into a SNAP-displacement sequence.
The size of the qudit Clifford group grows rapidly with dimension, however, quickly becoming prohibitively large.
Process tomography, meanwhile, is sensitive to SPAM errors, which poses a challenge when the gate errors themselves are smaller than SPAM errors.
%is challenging due to its sensitivity to SPAM errors.
We instead perform qubit randomized benchmarking using the spin cat encoding, where each logical Clifford gate is decomposed into a SNAP-displacement sequence \cite{heeres_gateset_2017}.
The average Clifford fidelity we recover is taken as a proxy for the overall fidelity of the displacement.
We begin our randomized benchmarking procedure with the SNAP-displacement sequence discussed above, which maps the ground state $\ket{0}$ onto the logical state $\ket{0_L}$.
We find that all logical Clifford unitaries involving logical $\hat{\sigma}_x$ or $\hat{\sigma}_y$ rotations can be implemented with two displacement pulses, while those corresponding to pure logical $\hat{\sigma}_z$ rotations require only a single phase update per drive, since the logical states are formed from disjoint sets of eigenstates.
We proceed in the typical fashion by sampling random sequences of Clifford gates of varying depth and appending the inverse Clifford operations to the end \cite{magesan_rb_2012}. 
We compile each Clifford gate down to a sequence of alternating SNAP and displacement gates.
The final step is to invert the encoding sequence such that the population of $\ket{0_L}$ is mapped back to $\ket{0}$ prior to readout.

We show selected randomized benchmarking data in Figure \ref{fig:rb}b for $d = 3$, $d = 5$, and $d = 8$.
The resulting average logical Clifford gate fidelities are $F_{\rm RB} = \{\num{\FRBthreelevel \pm \deltaFRBthreelevel}, \num{\FRBfivelevel \pm \deltaFRBfivelevel}, \num{\FRBeightlevel \pm \deltaFRBeightlevel} \}$ for $d = 3$, $d = 5$, and $d = 8$ respectively, obtained by fitting the exponential decay of the randomized benchmarking data.
While in the idealized case our control pulses implement the generator of spin displacements, we find that off-resonant drive terms neglected in the RWA can cause coherent errors such as phase shifts and leakage, which can have a significant impact on the displacement fidelity for short pulse durations.
We address this in several complementary ways.
To lowest order the phase offsets accumulated during the pulse can be compensated for by instantaneous phase corrections before and after the pulse.
We compute these phase corrections from numerical simulations of the unitary evolution of the system under the displacement drive.
We find that additional improvements to the fidelity are made by including small drive detunings and DRAG-like corrections, which correct the non-commuting phase errors not addressed by the instantaneous phase corrections and leakage, respectively.
We calculate these corrections numerically using an optimal control code \cite{machnes_goat_2018} \cite{supp}.
The optimal fidelities are achieved by utilizing all three types of corrections and by choosing displacement pulse lengths of $\qty{48}{\nano\second}$ for $d = 3$, $\qty{100}{\nano\second}$ for $d = 5$, and $\qty{140}{\nano\second}$ for $d = 8$ to balance coherent errors, which increase with pulse strength, and incoherent errors, which increase with pulse duration.
We further demonstrate the effects of our pulse corrections and the overall duration of the displacement pulse in Figure \ref{fig:rb}e for $d = 8$ by performing randomized benchmarking with displacement pulses of different durations and different combinations of pulse corrections.
As a guide to the eye we fit a phenomenological model in which the total infidelity is the sum of coherent and incoherent error terms: this model assumes that the coherent part decreases with the square of the pulse duration while the incoherent part increases linearly with pulse duration, and also assumes that each pulse correction type is subject to the same rate of incoherent error.
We see that the pulse corrections do indeed improve the fidelities, particularly for short pulse durations.
For longer pulses the incoherent error is dominant, so the fidelity curves converge to one another.
We further test the stability of our control scheme and the reliability with which our pulse corrections improve the fidelity by repeating the same randomized benchmarking experiment many times, as shown in Figure \ref{fig:rb}f.
The measured fidelities remain stable over roughly 82 hours, with the distributions of fidelities under each type of pulse correction remaining well-separated.

We note that our randomized benchmarking results require careful interpretation for several reasons.
First, the fidelity of the displacement over the full qudit Hilbert space is not necessarily equivalent to the fidelity within the logical subspace: it is conceivable that the displacement pulse is subject to errors affecting states orthogonal to the logical subspace, which would be suppressed in our results.
Second, randomized benchmarking can potentially fail to account for errors resulting in leakage out of the logical subspace because such errors can in general produce non-Markovian effects \cite{epstein_rb_limits_2014}.
Finally, the logical Clifford operations require differing numbers of displacement pulses.
We address these complications using a numerical model of our randomized benchmarking procedure in which randomly sampled completely positive trace-preserving maps are applied after each displacement \cite{heeres_gateset_2017}.
The average gate fidelity of this perturbed displacement is computed directly, which we take to be the underlying displacement fidelity $F_{\hat{D}}$ under a particular realization of the error channel.
Meanwhile, the perturbed displacement is used to construct the Clifford set, from which we sample sequences of varying length.
A fit to the resulting exponential decay is used to compute the average Clifford fidelity $F_{\rm RB}$.
Repeating this process for many randomizations and weights of the error map ultimately allows us to estimate the relationship between these two fidelities, shown in Figure \ref{fig:rb}c \cite{supp}.
We find that the simple parameter-free model $F_{\hat{D}} = (F_{\rm RB})^{1/N}$ represents the numerical data well, with $N = 5/3$ being the average number of displacement pulses per Clifford gate.
Using this relation and our Clifford fidelities from randomized benchmarking we obtain estimated displacement pulse fidelities of $F_{\hat{D}} = \{\num{\FDthreelevel \pm \deltaFDthreelevel}, \num{\FDfivelevel \pm \deltaFDfivelevel}, \num{\FDeightlevel \pm \deltaFDeightlevel} \}$ for $d = 3$, $d = 5$, and $d = 8$ respectively.

%(e.g. quantum fields \cite{ciavarella_trailhead_2021, gonzalez-cuadra_hardware_2022,gustafson_prospects_2021} or large spins \cite{senko_realization_2015, neeley_emulation_2009}). Specifically qudits are interesting because prime number qudits are great for QEC\cite{campbell_magic-state_2012}, qutrit encodings lead to asymptotic reductions in circuit complexity \cite{}and we can simulate systems with topological phases \cite{sompet_realizing_2022}.
\section{Discussion}

In this paper we have demonstrated a new method for controlling superconducting transmon qudits which combines concepts from bosonic computing platforms and high-dimensional spin systems to realize efficient, high-fidelity unitary control.
Our method establishes a single high-dimensional spin displacement operation as a primitive gate for qudit computation, which in combination with virtual phase rotations is universal over the single-qudit Hilbert space.
This, in combination with our single-shot projective readout of the full qudit state, takes full advantage of the qudit Hilbert space.
Our gate set is highly efficient as compared to typical decompositions of qudit unitaries, requiring $\mathcal{O}(d)$ physical pulses in our implementation to produce any desired unitary.
It furthermore does not in principle require optimal control methods, instead making use of the spectral separation of our qudit transitions to directly drive the generator of the spin displacements; this property makes it immediately applicable to transmon qudits of any dimension, and indeed to any qudit platform in which the transitions are well-separated in frequency.
We note that while in this work we apply optimal control in calculating the DRAG, detuning, and phase corrections, we expect that further theoretical work could yield analytic predictions for these parameters.
It may additionally be possible to find simple experimental methods for directly calibrating these corrections without the need for analytic predictions or accurate numerical simulations, which we leave for future work.

A simple model of the displacement pulse fidelity in terms of coherent and incoherent errors, presented in the Supplemental Material \cite{supp}, suggests that an increase in coherence time would have a considerable effect on the displacement fidelity.
We estimate that increasing our coherence times to be on par with state-of-the-art transmon devices \cite{wang_millisecond_2022} would increase our fidelities to $F_{\hat{D}}({\rm predicted}) = \{ 0.9993, 0.998, 0.996 \}$ for $d = 3$, $d = 5$, and $d = 8$ respectively, even in the absence of pulse corrections.
Our experimental results suggest that including pulse corrections could further decrease the infidelity by roughly a factor of 2.
Alternative device designs such as the inductively shunted transmon \cite{hassani_ist_2023} could eliminate the trade off between the number of usable qudit states and the anharmonicity, further suppressing coherent errors.

Qudit computation has the potential to significantly improve the capacity of current quantum computing architectures due to its intrinsic hardware efficiency, speeding up certain types of computations.
Our work represents a major step towards this goal, simultaneously utilizing a high-$E_J / E_C$ transmon design for many-level qudit computation and a SNAP-displacement decomposition for high-fidelity control.
Furthermore, we have demonstrated that logical qubits encoded into the transmon qudit Hilbert space can be controlled with high fidelity. Our multi-frequency control and readout techniques can be integrated with recent advances on qudit entangling gates \cite{goss_high-fidelity_2022,nguyen_empowering_2023} to explore the advantages of qudit technology.

In particular, there has recently been much interest in bosonic error-correcting codes in linear resonator modes \cite{gottesman_encoding_2001, ofek_extending_2016, grimm_stabilization_2020, sivak_real-time_2023, ni_beating_2023}, as well as error-correcting codes in high-dimensional spins \cite{gross_codes_2021, omanakuttan_cat_2024}.
In both cases a high-dimensional Hilbert space is used to implement a stabilizer code designed to correct the most physically relevant error processes for the system at hand. 
Our work motivates investigation into whether similar error correcting codes can be implemented in a transmon qudit, taking advantage of the high-fidelity control and readout enabled by our strong nonlinearity and multi-frequency drives while retaining the hardware efficiency inherent to bosonic and spin qudit codes.
In the context of quantum simulation, qudits have recently been proposed as natural building blocks for simulating lattice gauge theories \cite{ciavarella_trailhead_2021,illa_qu8its_2024,gonzalezcuadra_hardware_2022}, owing to their ability to naturally encode multiple degrees of freedom per site.
We ultimately expect that our control and readout scheme will facilitate more complex qudit computations and simulations than would otherwise be possible given current coherence times while remaining highly extensible and transferable between platforms.

\vspace{1em}

In the preparation of our manuscript, we became aware of two complementary works on distinct hardware platforms that also demonstrate large-spin $SU(2)$ dynamics and Schrodinger spin cats \cite{HAMSexpt,Yu2024}.

\section{Acknowledgements}

We would like to thank Shruti Puri and Daniel Weiss for fruitful discussions during the course of this work as well as feedback on the manuscript.
This material is based upon work supported by the Air Force Office of Scientific Research under award number FA9550-23-1-0121.
Devices were fabricated and provided by the Superconducting Qubits at Lincoln Laboratory (SQUILL) Foundry at MIT Lincoln Laboratory, with funding from the Laboratory for Physical Sciences (LPS) Qubit Collaboratory.

%%%%%%%%%%%%%%%%%%%%%%%%%%%%%%%%%%%%%%%%%%%%%%%%%%%%%%

% citations that only appear in the supplemental material
\nocite{shillito_ionization_2022}
\nocite{lloyd_universality_1995}
\nocite{scqubits_2021}
\nocite{scqubits_2022}
\nocite{qutip_2013}
\nocite{bruzda_random_2009}

\bibliography{Bibliography}
\bibliographystyle{naturemag}
\newpage
% % Modified from the APS RevTeX 4.2 template
% \documentclass[preprint,amsmath,amssymb,aps,biblatex]{revtex4-2}

% \usepackage{graphicx}% Include figure files
% \usepackage{dcolumn}% Align table columns on decimal point
% \usepackage{bm}% bold math
% \usepackage{lineno}
% \usepackage{braket}
% \usepackage{svg}
% \usepackage{siunitx}
% \usepackage{bbm}
% \usepackage{multirow}

% \setlength{\linenumbersep}{4pt}
% %\linenumbers

% \begin{document}
%%%%%%%%%%%%%%%%%%%%% numbers for results %%%%%%%%%%%%%%%%%%%%%

\onecolumngrid
\section{Supplemental Material: Multi-frequency control and measurement of a spin-7/2 system encoded in a transmon qudit}

\author{Elizabeth Champion}
    \thanks{These authors contributed equally to this work.}
    %\email{elizabeth.champion@rochester.edu}
\author{Zihao Wang}
    \thanks{These authors contributed equally to this work.}
\author{Rayleigh Parker}
\author{Machiel Blok}
    \email{machielblok@rochester.edu}
\affiliation{
 Department of Physics and Astronomy, University of Rochester, Rochester, NY 14627
}

\date{\today}

\maketitle

%\widetext

%%%%%%%%%% Prefix a "S" to all equations, figures, tables and reset the counter %%%%%%%%%%
\setcounter{equation}{0}
\setcounter{figure}{0}
\setcounter{table}{0}
\setcounter{page}{1}
\setcounter{section}{0}
\renewcommand{\thesection}{S-\Roman{section}}
\makeatletter
\renewcommand{\theequation}{S\arabic{equation}}
\renewcommand{\thefigure}{S\arabic{figure}}
\renewcommand{\thetable}{S\arabic{table}}
%\renewcommand{\bibnumfmt}[1]{[S#1]}
%\renewcommand{\citenumfont}[1]{S#1}
%%%%%%%%%% Prefix a "S" to all equations, figures, tables and reset the counter %%%%%%%%%%

\section{High $E_J/E_C$ transmon device}

Our implementation of spin displacements and qudit readout relies on having a physical device which can encode a many-level system with minimal decoherence while maintaining spectral separation of its transitions.
In this work we propose a superconducting transmon qudit with a large ratio of $E_J / E_C$.
The transmon Hamiltonian is
\begin{equation}
    \hat{H} = 4 E_C (\hat{n} - n_g)^2 - E_J \cos \hat{\phi}
\end{equation}
where $E_C$ is the capacitive energy, $E_J$ is the Josephson energy, $\hat{n}$ is the transmon charge operator, $\hat{\phi}$ is the superconducting phase operator, and $n_g$ is a gate charge \cite{blais_cqed_2021}.
The $\ket{0} \leftrightarrow \ket{1}$ transition frequency is approximately given by $\omega_{01} = \sqrt{8 E_J E_C} - E_C$, while the anharmonicity is $-E_C$.
A rough estimate of the number of levels contained in the cosine potential can be obtained by dividing the depth of the potential, $2 E_J$, by the plasma frequency $\sqrt{8 E_J E_C}$ \cite{shillito_ionization_2022}:
\begin{equation}
    n_{\rm levels} \approx \sqrt{\frac {E_J} {2 E_C}},
\end{equation}
so we see that we should expect more excited states to be confined in the cosine potential as $E_J / E_C$ is increased.

We must also consider the effects of change noise-induced dephasing in the higher excited states.
The charge dispersion of the transmon state $\ket{n}$ is given by \cite{koch_transmon_2007}
\begin{equation}
    \epsilon_n = (-1)^n E_C \frac {2^{4n + 5}} {n!} \sqrt{\frac {2} {\pi}} \left ( \frac {E_J} {2 E_C} \right )^{\frac {n} {2} + \frac {3} {4}} e^{-\sqrt{8 E_J / E_C}}.
\end{equation}
The key point is that while the charge dispersion increases exponentially with level, it decreases exponentially with $\sqrt{E_J / E_C}$.
To realize a qudit with 8 levels we target a ratio of $E_J / E_C \approx 270$.
We keep the transition frequency fixed to roughly $\qty{5}{\giga\hertz}$, which necessitates a decrease in the capacitive energy $E_C$ and therefore in the anharmonicity.
We find, however, that our anharmonicity is still large enough to permit high-fidelity control of the system.
In Table \ref{tab:device} we report the measured transition frequencies as well as calculated values for the underlying device parameters, and in Table \ref{tab:coherence} we report the measured coherence times.

\renewcommand{\arraystretch}{1.5}
\begin{table}[h!]
\begin{center}
\begin{tabular}{ c c c c c c c | c c c c }
    $f_{01}$ & $f_{12}$ & $f_{23}$ & $f_{34}$ & $f_{45}$ & $f_{56}$ & $f_{67}$ & $E_J$ & $E_C$ & $f_r$ & $g$ \\
    \hline
    4.896 & 4.782 & 4.664 & 4.539 & 4.407 & 4.267 & 4.116 & 29.09 & 0.108 & 6.410 & 0.028 \\
\end{tabular}
\caption{Measured transition frequencies and calculated $E_J$, $E_C$, bare resonator frequency $f_r$, and coupling strength $g$. All values are expressed in GHz.}
\label{tab:device}
\end{center}
\end{table}

\renewcommand{\arraystretch}{1.5}
\begin{table}[h!]
\begin{center}
\begin{tabular}{ c | c c c c c c c }
    & $\ket{0} \leftrightarrow \ket{1}$
    & $\ket{1} \leftrightarrow \ket{2}$
    & $\ket{2} \leftrightarrow \ket{3}$
    & $\ket{3} \leftrightarrow \ket{4}$
    & $\ket{4} \leftrightarrow \ket{5}$
    & $\ket{5} \leftrightarrow \ket{6}$
    & $\ket{6} \leftrightarrow \ket{7}$ \\
    \hline 
    $T_1$ (\SI{}{\micro\second}) & $46 \pm 7$ & $25 \pm 4$ & $26 \pm 3$ & $14 \pm 3$ & $16 \pm 2$ & $14 \pm 2$ & $13 \pm 2$ \\
    $T_2^*$ (\SI{}{\micro\second}) & $50 \pm 10$ & $24 \pm 6$ & $15 \pm 4$ & $15 \pm 5$ & $50 \pm 10$ & $30 \pm 10$ & $20 \pm 9$ \\
    $T_2$ (\SI{}{\micro\second}) & $51 \pm 7$ & $37 \pm 5$ & $34 \pm 3$ & $25 \pm 7$ & $44 \pm 6$ & $27 \pm 6$ & $23 \pm 6$ \\
\end{tabular}
\caption{Coherence times for each qudit transition. Each coherence time was measured 32 times over the course of 14 hours, and here we report the mean values and standard deviations.}
\label{tab:coherence}
\end{center}
\end{table}

\begin{figure}
    \centering
    \includegraphics[width=0.4\linewidth]{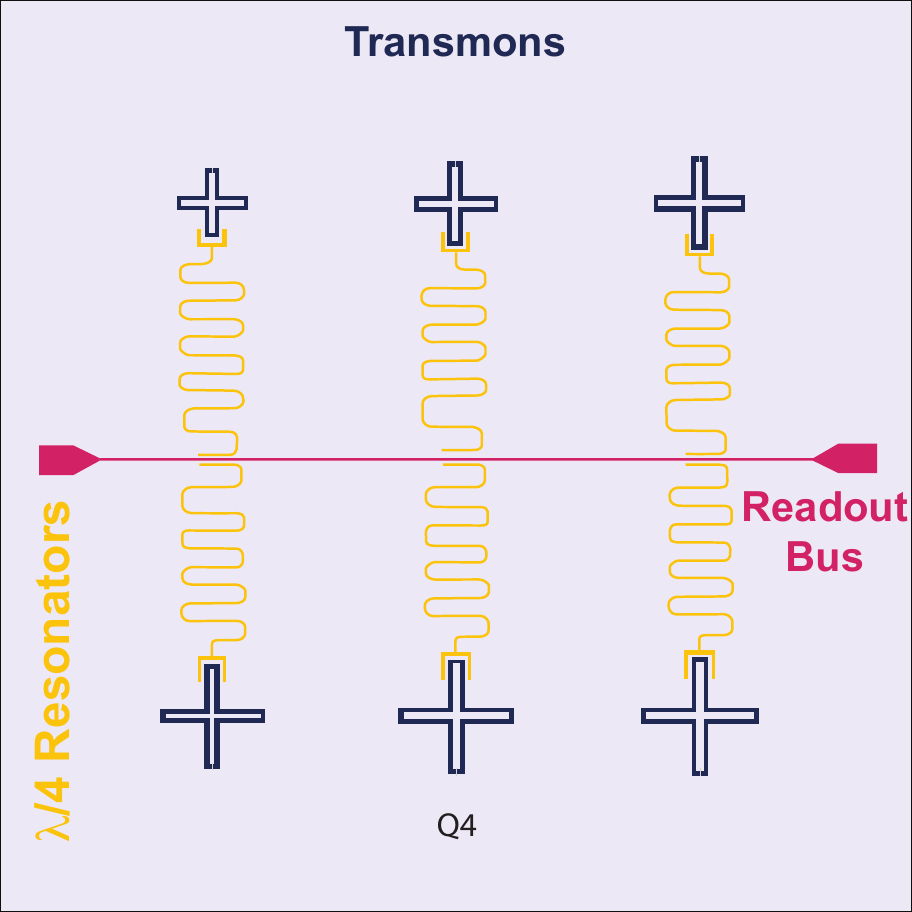}
    \caption{False-color image of the device design. All experiments in this work were carried out using the transmon labeled Q4.
    }
    \label{fig:device}
\end{figure}

We show a false-color image of our device design in Figure \ref{fig:device}; the varying sizes of the transmon capacitor pads reflect the range of values of $E_J / E_C$ chosen for this device.
All experiments in this work were performed using the transmon labeled Q4.
The device was fabricated in aluminum by the Superconducting Qubits at Lincoln Laboratory (SQUILL) Foundry at MIT Lincoln Laboratory.
\section{Single-shot dispersive qudit readout}

Single-shot readout of the transmon qudit state is enabled by the dispersive coupling between the transmon and the readout resonator.
In the lowest-order nonlinear approximation to the transmon potential the Hamiltonian of the transmon-resonator system is given by
\begin{equation}
    \hat{H} = \omega_r \hat{a}^\dag \hat{a} + \omega_q \hat{b}^\dag \hat{b} - \frac {E_C} {2} \hat{b}^{\dag 2} \hat{b}^2 + \sum_n \left ( \Lambda_n + \chi_n \hat{a}^\dag \hat{a} \right ) \ket{n} \bra{n}
\end{equation}
where $\hat{a}$ and $\hat{b}$ are the annihilation operators for the resonator and transmon respectively, $\omega_r$ and $\omega_q$ are the bare resonator and transmon frequencies, the $\Lambda_n$ are Lamb shifts, and the $\chi_n$ are dispersive shifts \cite{blais_cqed_2021, koch_transmon_2007}.
See Ref. \cite{blais_cqed_2021} for the detailed forms of the Lamb and dispersive shifts; the salient point is that the resonator has an effective frequency shift that is different for each transmon state.

We optimize our device design to have dispersive shifts comparable in size to the linewidth of the readout resonator, ultimately producing the transmon-state-dependent spectrum depicted in Figure 2b in the main text.
In this regime a single readout tone is able to distinguish several of the states, where there is appreciable overlap between the resonator spectra in those states.
In order to read out the full qudit state, we apply three simultaneous readout tones with frequencies chosen such that together they are able to unambiguously distinguish between all qudit states.
The signal transmitted through the readout line is demodulated at each of these frequencies and integrated to produce a 6-dimensional vector $(I_1, Q_1, I_2, Q_2, I_3, Q_3)$.

\begin{figure}
    \centering
    \includegraphics{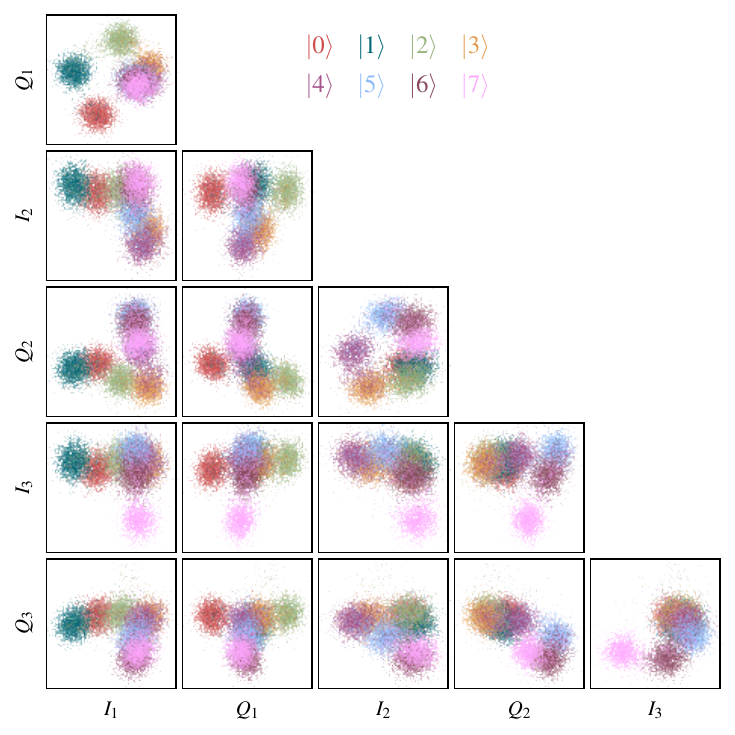}
    \caption{$IQ$ readout data for each qudit state.
    }
    \label{fig:readout}
\end{figure}

\renewcommand{\arraystretch}{1.5}
\begin{table}[h!]
\begin{center}
\begin{tabular}{ c  c | c c c c c c c c }
     & & \multicolumn{8}{c}{Prepared state} \\
    \multirow{8}{3em}[-3.5em]{\rotatebox{90}{Assigned state}} & & & & & & & & \\
    & & $\ket{0}$ & $\ket{1}$ & $\ket{2}$ & $\ket{3}$ & $\ket{4}$ & $\ket{5}$ & $\ket{6}$ & $\ket{7}$ \\
    %\cline{2-10}
    \hline
    & $\ket{0}$ & \textbf{0.99} & 0.04 & 0.01 & 0.01 & 0.01 & 0.01 & 0.01 & 0.01 \\
    & $\ket{1}$ & 0.01 & \textbf{0.96} & 0.07 & 0.01 & 0.01 & 0 & 0 & 0.01 \\
    & $\ket{2}$ & 0 & 0 & \textbf{0.89} & 0.05 & 0.01 & 0 & 0 & 0 \\
    & $\ket{3}$ & 0 & 0 & 0.03 & \textbf{0.92} & 0.18 & 0.02 & 0.01 & 0.01 \\
    & $\ket{4}$ & 0 & 0 & 0.01 & 0.01 & \textbf{0.79} & 0.09 & 0.01 & 0.01 \\
    & $\ket{5}$ & 0 & 0 & 0 & 0 & 0 & \textbf{0.86} & 0.10 & 0.08 \\
    & $\ket{6}$ & 0 & 0 & 0 & 0 & 0 & 0.01 & \textbf{0.86} & 0.10 \\
    & $\ket{7}$ & 0 & 0 & 0 & 0 & 0 & 0 & 0.01 & \textbf{0.79} \\
\end{tabular}
\caption{Probabilities that a prepared state $\ket{n}$ will be classified as $\ket{m}$.} \label{tab:assignment}
\end{center}
\end{table}

We represent the resulting distributions in this 6-dimensional space in Figure \ref{fig:readout}, where points are colored according to the state that was prepared.
We calibrate state classification by fitting a 6-dimensional Gaussian mixture model to this data with each Gaussian component corresponding to one of the qudit states.
The resulting assignment probabilities are shown in Table \ref{tab:assignment}, yielding an average assignment fidelity of $(88.3 \pm 0.2) \%$, where the uncertainty is calculated from the $N = 5000$ experimental shots and assuming binomial statistics.
In practice these probabilities are used to calculate a correction matrix, which is applied to the populations measured in each of the experiments in the main text.
\section{Lie groups and Lie algebras}

We begin our discussion of qudit displacements, phase spaces, and Wigner functions with a pedagogical overview of Lie groups and their application to quantum systems with dynamical symmetries.
Central to the idea of a phase space is the concept of a dynamical symmetry group.
Recall that a group $G$ is a collection of elements $g \in G$ and a binary operation between these group elements (which we denote by multiplication) having the following properties:
\begin{enumerate}
    \item The group operation is associative: for any $g_a, g_b, g_c \in G$, we have $(g_a g_b) g_c = g_a (g_b g_c)$.
    \item There exists a unique identity element $g_e \in G$ such that for any $g_a \in G$, $g_e g_a = g_a g_e = g_a$.
    \item Every $g \in G$ has a unique inverse element, denoted by $g^{-1}$, such that $g^{-1} g = g g^{-1} = g_e$.
\end{enumerate}
Any collection satisfying these properties is a group, irrespective of the underlying nature of its elements.
If two groups share the same structure -- in the sense that there is a bijective mapping between their elements under which they have the same multiplication table -- then we say that they are isomorphic, meaning that they are, in some sense, the same group.

Groups frequently arise in physics when a system is symmetric under some set of transformations.
We focus here on the case of a Lie group, which applies to continuous symmetries, such as rotational symmetry.
A $d$-dimensional Lie group is a group that can be parameterized by $d$ continuously-varying, real-valued parameters $x_1, x_2, \ldots, x_d$, which we collectively denote by a vector $\mathbf{x}$.
That is, it is a smooth manifold that also has a group structure.
The group elements, $g$, are functions of the parameters: $g = G(\mathbf{x})$.

It is often convenient to construct a Lie group in terms of its Lie algebra, a set of algebraic objects which generate the Lie group through exponentiation.
The generators are related to the group elements themselves in that they are the infinitesimal group elements.
They represent a linearization of the action of group elements; concretely, they are given by 
\begin{equation}
    X_j = i \left ( \frac {\partial g(\mathbf{x})} {\partial x_j} \right )_{\mathbf{x} = 0}.
\end{equation}
We therefore see that the number of generators is equal to $d$, the number of dimensions.
The generators of a Lie group form a vector space, meaning that linear combinations of them are also generators, and therefore generate group elements.
They obey the commutation relations 
\begin{equation}
    [X_i, X_j] = i f_{ijk} X_k,
\end{equation}
where the $f_{ijk}$ are known as structure constants.
When we think of a specific, concrete realization of a Lie group or Lie algebra -- for example, as operators acting on a Hilbert space -- we are invoking the idea of a group representation.
We denote the elements of a group's representation as $D$:
\begin{equation}
    D[g(\mathbf{x})] \equiv D(\mathbf{x}).
\end{equation}
The relationship between the representation of a group and the representation of its generators is given by an exponential map,
\begin{equation}
    D[g(\mathbf{x})] = e^{-i \sum_j x_j X_j}.
\end{equation}
\section{The harmonic oscillator}

It is helpful to begin with a description of the standard coherent states of a harmonic oscillator and its phase space representation.
We refer the reader to Refs. \cite{perelomov_generalized_1977, davis_wigner_negativity_2021, stratonovich_distributions_1957, heiss_discrete_2000} for more comprehensive discussions of this and spin displacements.
The harmonic oscillator is conveniently described by the bosonic creation and annihilation operators $\hat{a}^\dag$ and $\hat{a}$, which together with the identity operator $\hat{\mathbbm{1}}$ obey the Heisenberg commutation relations:
\begin{equation}
    [\hat{a}, \hat{a}^\dag] = \hat{\mathbbm{1}},
\end{equation}
\begin{equation}
    [\hat{a}, \hat{\mathbbm{1}}] = [\hat{a}^\dag, \hat{\mathbbm{1}}] = 0.
\end{equation}
These operators therefore define a Lie algebra, having a general element of the form 
\begin{equation}
    t \hat{\mathbbm{1}} - i (\alpha \hat{a}^\dag - \alpha^* \hat{a})
\end{equation}
for a real number $t$ and complex number $\alpha$ (such a parameterization consists of three independent real values, reflecting the dimension of the algebra, while enforcing Hermiticity).
It follows that the operators 
\begin{equation}
    \hat{T}(t, \alpha) = e^{it} e^{\alpha \hat{a}^\dag - \alpha^* \hat{a}}
\end{equation}
are a representation of a Lie group, known as the Heisenberg-Weyl group $H_3$.
This Lie group represents the dynamical symmetry group of the harmonic oscillator when it is acted upon by an external drive.
We therefore identify
\begin{equation}
    \hat{D}(\alpha) \equiv e^{\alpha \hat{a}^\dag - \alpha^* \hat{a}}
\end{equation}
as the harmonic oscillator displacement operator; the multiplication of two such operators yields
\begin{equation}
    \hat{D}(\alpha) \hat{D}(\beta) = e^{i \text{Im}(\alpha \beta^*)} \hat{D}(\alpha + \beta).
\end{equation}
A set of coherent states can, in general, be constructed by applying these displacement operators to some fixed state in the Hilbert space.
Making the typical choice of $\ket{0}$ as our fixed state, we have
\begin{equation}
    \ket{\alpha} \equiv \hat{D}(\alpha) \ket{0}.
\end{equation}
This choice of fixed state yields a set of coherent states equivalent to those of the typical formulation, defined as eigenstates of the annihilation operator:
\begin{equation}
    \hat{a} \ket{\alpha} = \alpha \ket{\alpha}.
\end{equation}

We now turn our attention to the phase space representation of a harmonic oscillator state -- in particular, the Wigner function.
For any quantum system, the Wigner function can be defined as 
\begin{equation}
    W(\alpha) = \text{Tr}[\hat{\rho} \hat{\Delta}(\alpha)]
\end{equation}
where $\hat{\rho}$ is the density matrix and $\hat{\Delta}(\alpha)$ is a special operator-valued function over the phase space, called the \textit{kernel}.
For a harmonic oscillator the kernel may be written in terms of a displacement in phase space and the parity operator as 
\begin{equation}
    \hat{\Delta}(\alpha) = 2 \hat{D}(\alpha) \hat{\Pi} \hat{D}^\dag(\alpha);
\end{equation}
substituting this into the definition of the Wigner function and exploiting the cyclic property of traces yields the familiar equation
\begin{equation}
    W(\alpha) = 2 \text{Tr}[\hat{D}^\dag(\alpha) \hat{\rho} \hat{D}(\alpha) \hat{\Pi}] = 2 \text{Tr}[\hat{D}(-\alpha) \hat{\rho} \hat{D}(\alpha) \hat{\Pi}].
\end{equation}
\section{Spin systems}

While the results in the previous section are specific to the case of a harmonic oscillator, the framework summarized here is far more general, allowing one to construct the Wigner function for a variety of systems.
The mapping $W(\alpha) = \text{Tr}[\hat{\Delta}(\alpha) \hat{A}]$ from operators $\hat{A}$ to functions on a phase space is mediated by the self-dual kernel $\hat{\Delta}(\alpha)$, which can be derived for a specific system from a set of conditions known as the Stratonovich-Weyl postulates \cite{stratonovich_distributions_1957, heiss_discrete_2000, davis_wigner_negativity_2021}.
The process for finding such a kernel is intimately related to the displacement operation, and therefore to the coherent states of the system.
In its most general form this calculation is very involved, so here we only discuss the construction of a spin displacement operator and then quote a result for the kernel in terms of the displacement and a generalization of a parity operator.

The Hilbert space of a spin-$j$ system is spanned by a basis of $2j + 1$ states, which we label $\ket{j,m}$ for $-j \leq m \leq j$.
These are the eigenstates of the angular momentum operator $\hat{J}_z$.
The dynamical symmetry group of a spin is $SU(2)$, and the Lie algebra generating this group is spanned by the basis $\{\hat{J}_x, \hat{J}_y, \hat{J}_z \}$,
\begin{equation}
    [\hat{J}_x, \hat{J}_y] = i \hat{J}_z, \quad
    [\hat{J}_y, \hat{J}_z] = i \hat{J}_x, \quad
    [\hat{J}_z, \hat{J}_x] = i \hat{J}_y.
\end{equation}
One can equivalently work in the basis $\{\hat{J}_+, \hat{J}_-, \hat{J}_z \}$, where 
\begin{equation}
    \hat{J}_\pm = \hat{J}_x \pm i \hat{J}_y 
\end{equation}
are ladder operators for the spin state:
\begin{equation}
    \hat{J}_\pm \ket{j,m} = \sqrt{j(j + 1) - m(m \pm 1)} \ket{j,m \pm 1}.
\end{equation}
In this basis we have 
\begin{equation}
    [\hat{J}_z, \hat{J}_\pm] = \pm \hat{J}_\pm, \quad [\hat{J}_+, \hat{J}_-] = 2 \hat{J}_z.
\end{equation}
A general group element can be parameterized as 
\begin{equation}
    \hat{g}(\psi, \theta, \phi) = e^{-i \phi \hat{J}_z} e^{-i \theta \hat{J}_y} e^{-i \psi \hat{J}_z}
\end{equation}
where $\phi, \theta, \psi \in \mathbbm{R}$ can be understood as Euler angles.
We therefore see that the phase space associated with a spin is the surface of a sphere, $S^2$.

Much like the case of a harmonic oscillator, the coherent states of a spin are the states generated by applying our group element representation, $\hat{g}$, to a fixed state, which we choose to be $\ket{j, j}$.
We see that the effect of the $e^{-i \psi \hat{J}_z}$ term is to add a trivial phase factor to our states, and so we take $\psi = 0$ and define our displacement operator:
\begin{equation}
    \hat{D}_j(\theta, \phi) = e^{-i \phi \hat{J}_z} e^{-i \theta \hat{J}_y}.
\end{equation}
It can be shown \cite{perelomov_generalized_1977} that this is equivalent to
\begin{equation}
    \hat{D}_j(\alpha) = e^{\alpha \hat{J}_+ - \alpha^* \hat{J}_-}
\end{equation}
where 
\begin{equation}
    \alpha = -\frac {\theta} {2} e^{-i \phi}.
\end{equation}
The coherent states of a spin, then, are 
\begin{equation}
    \ket{j,\alpha} = \hat{D}_j(\alpha) \ket{j,j}
\end{equation}
for $\alpha$ lying in a disk of radius $\frac {\pi} {2}$ in the complex plane.

\begin{figure}
    \centering
    \includegraphics{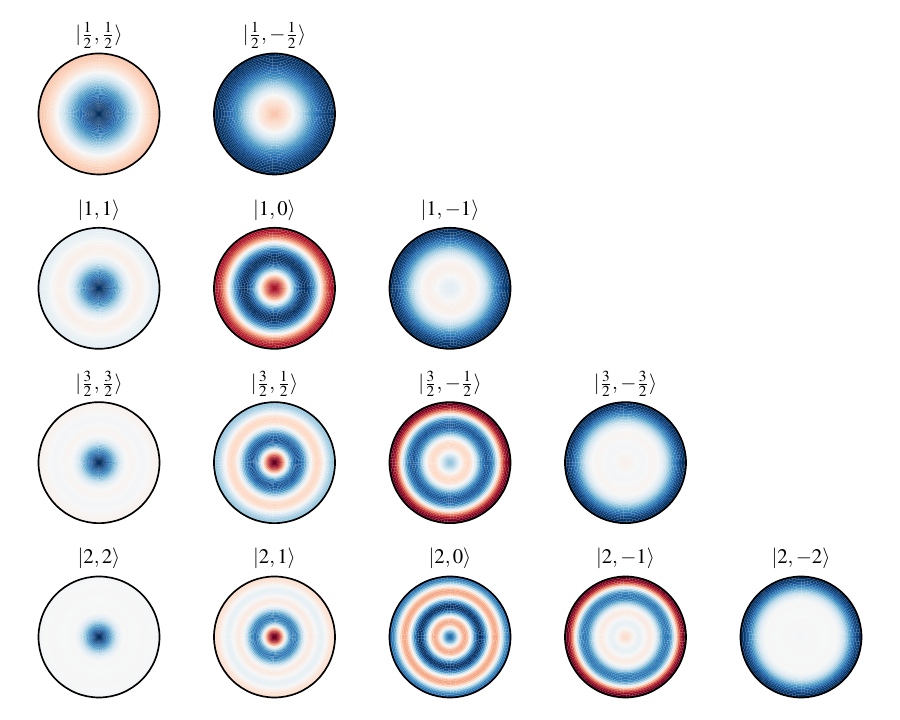}
    \caption{Wigner functions for $\hat{J}_z$ eigenstates.
    The spherical phase space is flattened to a circle where $\theta$ and $\phi$ are the radial and angular coordinates, respectively. The entire outer circumference of the plot, then, represents the south pole of the phase space. The bottom row corresponds to Figure 3b in the main text.
    }
    \label{fig:Jz_wigner}
\end{figure}

As stated above, the Wigner function for a given system is a mapping between operators on the Hilbert space and functions on the phase space.
In the present case of a spin-$j$ system, this mapping can be shown to be mediated by the kernel \cite{heiss_discrete_2000, davis_wigner_negativity_2021}
\begin{equation}
    \hat{\Delta}(\alpha) = \sum_{m=-j}^j \sum_{l=0}^{2j} \frac {2l + 1} {2j + 1} \cg{j}{m}{l}{0}{j}{m} \hat{D}(\alpha) \ket{j,m} \bra{j,m} \hat{D}^\dag (\alpha)
\end{equation}
where 
\[
    \cg{j_1}{m_1}{j_2}{m_2}{J}{M}
\]
is a Clebsch-Gordan coefficient.
Therefore, defining
\begin{equation}
    2 \hat{\Pi}_j = \sum_{m=-j}^j \sum_{l=0}^{2j} \frac {2l + 1} {2j + 1} \cg{j}{m}{l}{0}{j}{m} \ket{j,m} \bra{j,m},
\end{equation}
we see that the kernel can be written as 
\begin{equation}
    \hat{\Delta}(\alpha) = 2 \hat{D}(\alpha) \hat{\Pi}_j \hat{D}^\dag (\alpha),
\end{equation}
in direct analogy to the harmonic oscillator.
Note that the $\hat{\Pi}_j$ is not exactly a parity operator in the typical sense, but it is diagonal in the basis of $\hat{J}_z$ eigenstates.
\section{Unitary qudit control with displacement and SNAP gates}

The only case for which the spin displacement unitary alone provides universal control over the qudit Hilbert space is $d = 2$.
In this case the displacement operator reduces to rotations on the Bloch sphere, because the angular momentum operators coincide with the Pauli operators.
For $d > 2$, on the other hand, the angular momentum operators alone do not form a basis for all $d \times d$ Hermitian matrices, and therefore there are some unitaries which cannot be realized by rotations of the angular momentum alone.

In bosonic systems universal control is often realized by a decomposition of the target unitary into displacement gates and selective number-dependent arbitrary phase (SNAP) gates \cite{heeres_snap_2015, krastanov_snap_2015, kudra_negative_2022, ma_control_2021}.
We show here that the same scheme can be utilized for transmon qudits with spin displacements, with the added benefit that our SNAP gates are implemented as instantaneous virtual rotations.

A SNAP gate is parameterized in terms of a $d$-dimensional vector of phases that are applied to each eigenstate:
\begin{equation}
    \hat{S}(\vec{\phi}) = \prod_{n=0}^{d-1} e^{i \phi_n \ket{n}\bra{n}}.
\end{equation}
In this decomposition an arbitrary unitary is decomposed into a sequence of $N + 1$ SNAP gates interleaved with $N$ displacement gates:
\begin{equation}
    \hat{U} \big ( \vec{\phi}^{(0)}, \theta^{(1)}, \vec{\phi}^{(1)} \ldots \theta^{(N)}, \vec{\phi}^{(N)} \big ) = \hat{S}(\vec{\phi}^{(N)}) \hat{D}(\theta^{(N)}) \ldots \hat{S}(\vec{\phi}^{(1)}) \hat{D}(\theta^{(1)}) \hat{S}(\vec{\phi}^{(0)}).
\end{equation}
Note that we parameterize the displacement gates in terms of the amplitude, $\theta$, setting the phase of the displacement to zero.
This is because the displacement phase can be absorbed into the adjacent SNAP gates.
In practice we also take advantage of the fact that any rotation in $SU(2)$ can be decomposed into two rotations by $\pi/2$ interleaved with phase rotations, thereby replacing each displacement layer with two displacements by $\theta = \pi / 2$ separated by a rotation about $\hat{J}_z$, which is itself a particular case of a SNAP gate.
This potentially doubles the number of required pulses, but has the benefit that we only need to calibrate a single displacement rather than a continuous set of them.

\begin{figure}
    \centering
    \includegraphics{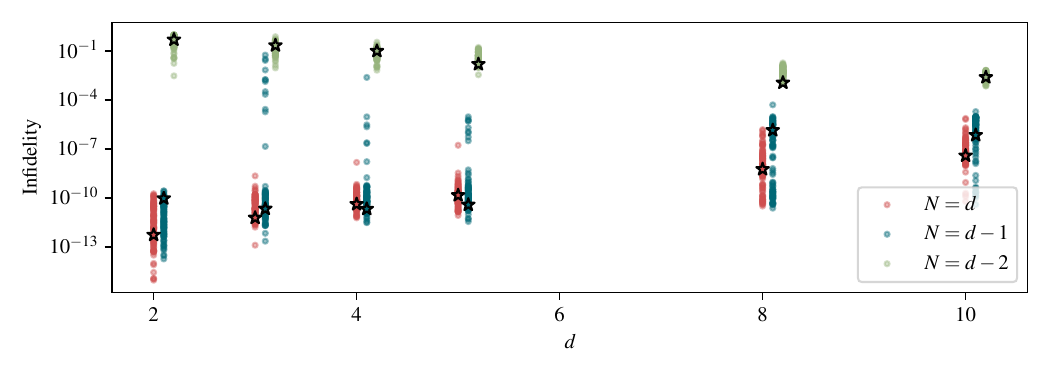}
    \caption{Numerical justification of the $\mathcal{O}(d)$ scaling of the SNAP-displacement decomposition.
    The circular markers correspond to each of the 100 Haar-random unitaries, while the stars correspond to the qudit Hadamard gate $\hat{H}_d$.
    Note that we have applied small horizontal offsets to the data points of differing decomposition depths to aid readability.
    }
    \label{fig:decomposition}
\end{figure}

\subsection{Proof of universality}

Here we show that the SNAP-displacement decomposition is universal over the qudit Hilbert space given large enough $N$, and provide numerical evidence below that $N$ scales as $\mathcal{O}(d)$.
Consider a set of gates generated through the exponentiation of $k$ different Hermitian operators $\hat{H}_1, \hat{H}_2, \ldots \hat{H}_k$ acting on a $d$-dimensional Hilbert space: $\hat{U}_m = e^{i t \hat{H}_m}$, where $t$ can be continuously varied.
The set of unitary operations that this gate set can realize consists of all $\hat{U} = e^{i \hat{L}}$, where $\hat{L}$ is any element of the algebra of Hermitian operators generated by taking commutators and repeated commutators of the operators $\hat{H}_m$; for example, $i[\hat{H}_1, \hat{H}_2]$, $[\hat{H}_1, [\hat{H}_1, \hat{H}_3]]$, and so on \cite{lloyd_universality_1995}.
If this algebra is a complete basis of $d \times d$ Hermitian matrices, then the gate set is capable of producing any rotation in $SU(d)$.
That is, it is universal.

With the phase of the displacement drive fixed to zero, the generator of displacements is the operator $\hat{J}_y$.
To understand the effect of interleaving displacements with SNAP gates, we first need a basis for the generators of the SNAP gates.
We are free to choose any complete basis of diagonal, real-valued $d \times d$ matrices, and we will see that the following operators are a particularly convenient choice:
\begin{equation}
    \hat{Q}_n = \sum_{q = 0}^{n} \ket{q} \bra{q}
\end{equation}
for $0 \leq n \leq d-2$.
Choosing one of these $\hat{Q}_n$ and taking its commutator with $\hat{J}_y$ yields
\begin{equation}
    i [\hat{J}_y, \hat{Q}_n] = \frac {1} {2} \sqrt{(n + 1) (d - (n + 1))} \bigg ( \ket{n} \bra{n + 1} + \ket{n + 1} \bra{n} \bigg ).
\end{equation}
We therefore see that by choosing $n$ we can generate a $\hat{\sigma}_x$-like operator in the $\{\ket{n}, \ket{n+1}\}$ subspace using a single commutator.
Treating each of these subspaces as a qubit, we see that combining these $\hat{\sigma}_x$ rotations with virtual phase rotations produces $\hat{\sigma}_y$ rotations in the subspace in question.
Finally, it is well known that this set of Pauli-like rotations between adjacent levels of a qudit is sufficient to generate any rotation in $SU(d)$ \cite{liu_performing_2023, nguyen_empowering_2023, deGuise_factorization_2018}, so we conclude that our gate set is universal.
We note that our proof is entirely analogous to that given in Ref. \cite{krastanov_snap_2015} for the case of a harmonic oscillator displacement.

\subsection{Linear scaling of the SNAP-displacement decomposition depth with dimension}

While the above proof shows that the SNAP-displacement decomposition is universal, it does not place any bounds on the number of layers required to decompose a given unitary with a satisfactory fidelity.
We can place a lower bound on the required depth by noting that the number of independent parameters in a $d$-dimensional unitary is $d^2 - 1$.
Assuming the displacement phase is fixed to zero, a displacement operator contains a single independent parameter, while a SNAP gate contains $d - 1$.
A SNAP-displacement decomposition of depth $N$ therefore has $N + (N + 1)(d - 1)$ parameters, so we expect to require a minimum of $N = d - 1$ SNAP-displacement layers to realize general unitaries.
We also note that it is possible to use fewer layers in cases where the desired unitary is closely related to the displacement operator itself, as is the case for the logical qubit Clifford rotations discussed in the main text.

The non-commuting nature of our SNAP and displacement gates makes a rigorous proof of universality for a fixed depth $N$ difficult, and it is unclear whether such a proof can be found.
We instead provide numerical evidence that the number of layers scales as $\mathcal{O}(d)$.
For a given qudit dimension $d$, we sample $100$ Haar-random $d \times d$ unitary matrices and attempt to find SNAP-displacement parameters producing each.
We use SNAP-displacement depths of $N = d - 2$, $N = d - 1$, and $N = d$ for qudit dimensions between $d = 2$ and $d = 10$ and show the results in Figure \ref{fig:decomposition}.
The optimizer fails to find satisfactory decomposition parameters for $N = d - 2$, as expected from the number of independent parameters.
For $N = d - 1$, meanwhile, the fidelities are very close to $1$ in all but a few cases.
It is unclear whether the points with unsatisfactory fidelities are due to the numerical optimizer becoming stuck in a local minimum or if the decomposition is fundamentally incapable of producing those unitaries.
For $N = d$, on the other hand, the decomposition is able to faithfully produce the desired unitary in every case.
We attribute the slight upwards trend in infidelity with dimension to the difficulty of performing numerical optimization in such a large parameter space.
We also show decomposition fidelities for the case of the qudit Hadamard gate $\hat{H}_d$, which follow the same overall trends as the random unitaries.
\section{DRAG, detuning, and phase corrections}

While our displacement drive ideally implements the generator of spin displacements, we find that off-resonant drives neglected in the RWA can cause phase errors and leakage.
These effects can be mitigated by increasing the pulse duration, thereby decreasing the drive amplitudes.
However, incoherent errors due to relaxation and dephasing limit the pulse duration.

We employ DRAG, detuning, and phase corrections to mitigate these coherent errors, permitting shorter pulses with smaller incoherent errors.
We calculate these corrections from numerical simulations of our system under the displacement drive.
We first compute the propagator of the system $\hat{U}(t)$ given the system Hamiltonian $\hat{H}(t) = \hat{H}_0 + \hat{H}_d(t)$ for transmon Hamiltonian $\hat{H}_0$ and time-dependent drive Hamiltonian $\hat{H}_d(t)$:
\begin{equation}
    \frac {d} {dt} \hat{U}(t) = -i \hat{H}(t) \hat{U}(t).
\end{equation}
This simulation is carried out in the lab frame without making the rotating wave approximation, using the transmon charge operator as the drive operator, starting at $t = 0$ and ending at the pulse duration $t = T$.
Following the simulation this propagator is transformed into the transmon frame: $\tilde{U}(T) = e^{i \hat{H}_0 t} \hat{U}(T)$.
Its fidelity with respect to the ideal displacement is then
\begin{equation}
    F = \frac {1} {d^2} \left | {\rm Tr} \left [ \hat{D}^\dag(\pi/2) \tilde{U}(T) \right ] \right |^2.
\end{equation}
Instantaneous phase corrections take the form of SNAP gates applied immediately before and after the pulse.,
\begin{equation}
    \hat{U}_{\rm corr}(T) = \hat{S}(\vec{\phi}_{\rm post}) \hat{U}(T) \hat{S}(\vec{\phi}_{\rm pre}).
\end{equation}
We compute the propagator once, then use numerical optimization to find the SNAP gate phases that maximize the fidelity with respect to the ideal displacement.

DRAG and detuning corrections, meanwhile, require optimizing the properties of the pulse waveform itself.
To this end we employ gradient-based optimal control.
Given the time-dependent propagator $\hat{U}(t)$, its derivative with respect to a control parameter $x$ is given by \cite{machnes_goat_2018}
\begin{equation}
    \frac {d} {dt} \hat{U}'_x(t) = -i \left ( \hat{H}'_x(t) \hat{U}(t) + \hat{H}(t) \hat{U}'_x(t) \right ).
\end{equation}
This differential equation is solved for each control parameter (the DRAG weights and drive detunings), then the gradient of the fidelity is computed using the chain rule.
The fidelity and its gradient are passed to a numerical optimizer, which terminates when the gradient becomes sufficiently small.
We note that the our transmon Hamiltonian is modeled using scqubits \cite{scqubits_2021, scqubits_2022}, and the simulations are in part carried out using QuTiP \cite{qutip_2013}.
\section{Validating randomized benchmarking fidelities}

The randomized benchmarking protocol utilized in the main text is generally only applicable under certain assumptions regarding the error processes affecting the logical state. In particular, leakage out of the logical subspace must be handled with care. Furthermore, even in the absence of leakage the fidelity estimated by randomized benchmarking is only valid within the logical subspace itself, making generalization to the overall displacement fidelity difficult.

We address this challenge here using a numerical model for the effect of noise on the displacement unitary and the resulting Clifford operations. We begin with a general quantum channel describing a completely positive trace-preserving (CPTP) perturbation to the displacement:
\begin{equation}
    \mathcal{D}(\hat{\rho}) = (1 - q) \hat{D} \hat{\rho} \hat{D}^\dag + q \mathcal{A}(\hat{D} \hat{\rho} \hat{D}^\dag).
\end{equation}
The error weight $q$, $0 \leq q \leq 1$, describes the strength of the error channel, with $q = 0$ corresponding to a perfect displacement operation. The superoperator $\mathcal{A}$ is taken to be a random CPTP map; thus in the case where $q = 1$ the system undergoes a completely random evolution instead of the displacement. We draw $\mathcal{A}$ from an ensemble of CPTP maps using QuTiP \cite{qutip_2013}, which follows the protocol described in Ref. \cite{bruzda_random_2009}. One can view the channel $\mathcal{D}$ as a Kraus decomposition in which we have the Kraus operator $\sqrt{1 - q} \hat{D}$ corresponding to a perfect displacement operation and Kraus operators $\sqrt{q} \hat{A}_i \hat{D}$ for each operator $\hat{A}_i$ in the Kraus decomposition of $\mathcal{A}$ corresponding to the error channel's effect on the ideal displacement.

For a particular realization of the perturbed displacement map, we begin by computing the average gate fidelity of this map with respect to the ideal displacement operation. We take this quantity to be the underlying fidelity of the displacement. To relate this number to a randomized benchmarking result, we compute the map corresponding to each logical Clifford operation using the perturbed displacement. We then simulate randomized benchmarking of these Clifford operations, extracting a Clifford fidelity from a fit to the exponential decay. By repeating this procedure for many values of $q$ and randomizations of the error channel we extract an approximate relationship between the measured Clifford fidelity and the underlying displacement fidelity, shown in Figure 4c in the main text.

\section{Outlook}

In this section we propose a simplified model for the displacement pulse errors, and using this model, present an estimate for realistically achievable fidelities given state-of-the-art transmon devices.
We begin with the model presented in the main text, where the total error is a sum of coherent and incoherent parts which vary independently depending on the pulse properties and the device coherence times:
\begin{equation}
    \mathcal{E} = \mathcal{E}_{\rm coh} + \mathcal{E}_{\rm inc}.
\end{equation}
We assume that the incoherent part of the error is proportional to the length of the pulse, with the proportionality constant being the average decay rate of each qudit transition: for a pulse duration $T$
\begin{equation}
    \mathcal{E}_{\rm inc} = \frac {T} {\tau_{\rm inc}},
\end{equation}
where for a $\ket{0} \leftrightarrow \ket{1}$ decay time $T_{01}$,
\begin{equation}
    \frac {1} {\tau_{\rm inc}} = \frac {1} {d-1} \sum_{n = 1}^{d - 1} \frac {n} {T_{01}} = \frac {d} {2 T_{01}},
\end{equation}
assuming a linear scaling of the decay rate with excitation number.
Here we want to consider transmons of varying qubit frequency, in which case the natural quantity to consider is the $Q$ factor: $Q = T_{01} f_{01}$.
We note that while $Q$ is not necessarily fixed as one varies qubit frequency, with its detailed behavior being dependent on the particular noise sources at hand, it is still the relevant quantity to use to relate the coherence properties of transmons across frequencies.
With this in mind we have
\begin{equation}
    \mathcal{E}_{\rm inc} = \frac {f_{01} d} {2 Q} T.
\end{equation}

We primarily attribute the coherent errors, meanwhile, to phase errors and leakage resulting from the off-resonant drive terms neglected in the RWA.
This suggests that the coherent error should scale proportionally with the square of the drive strength and inversely with the square of the anharmonicity.
In the following we assume a fixed ratio $E_J / E_C = 270$, chosen to ensure a large enough number of confined states with sufficiently low charge dispersion to permit the qudit operations considered here.
Using the approximate qubit frequency $f_{01} = \sqrt{8 E_J E_C} - E_C$ we can write
\begin{equation}
    E_C = \frac {f_{01}} {\sqrt{8 \frac {E_J} {E_C}} - 1}.
\end{equation}
Noting that the anharmonicity is equal in magnitude to $E_C$, then,
\begin{equation}
    \mathcal{E}_{\rm coh} = \frac {A(E_J, E_C, d)} {E_C^2 T^2}
\end{equation}
where $A(E_J, E_C, d)$ is a dimensionless phenomenological parameter capturing the time-independent scale of coherent errors; because $E_J/E_C$ is fixed we can equivalently parameterize it as $A(f_{01}, d)$.
Substituting our equation for $E_C$ we have
\begin{equation}
    \mathcal{E}_{\rm coh} = \left ( \sqrt{8 \frac {E_J} {E_C}} - 1 \right )^2 \frac {A(f_{01}, d)} {f_{01}^2 T^2}.
\end{equation}

\begin{figure}
    \centering
    \includegraphics{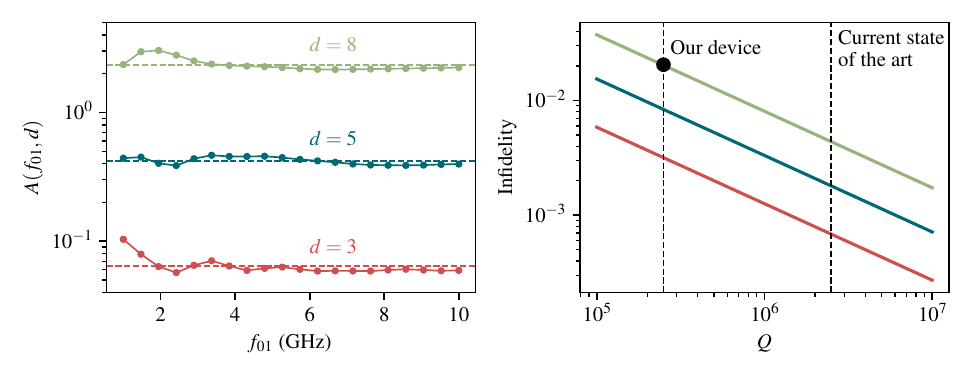}
    \caption{Modeling achievable displacement pulse fidelities.
    }
    \label{fig:outlook}
\end{figure}

Solving for the optimal pulse duration we find
\begin{equation}
    T_{\rm opt} = \frac {1} {f_{01}} \left [ 4 \left ( \sqrt{8 \frac {E_J} {E_C}} - 1 \right )^2 \frac {Q A(f_{01}, d)} {d} \right ]^{1/3}.
\end{equation}
We note that upon substituting this into the equations for $\mathcal{E}_{\rm coh}$ and $\mathcal{E}_{\rm inc}$, the factor of $f_{01}$ cancels out everywhere except in the argument to $A(f_{01}, d)$.
In order to characterize this dependence we perform a series of numerical simulations at differing transmon frequencies and extract $A(f_{01}, d)$ for each, showing the results in Figure \ref{fig:outlook} (left).
We find that to a good approximation $A(f_{01}, d)$ is independent of $f_{01}$, and so we are justified in considering it to be a function of $d$ only.
We ultimately find, therefore, that the minimum infidelity is given by
\begin{equation}
    \mathcal{E}_{\rm min} = \frac {3} {2} \left ( \frac {A(d)} {2} \right )^{1/3} \left ( \frac {d} {Q} \right )^{2/3} \left ( \sqrt{8 \frac {E_J} {E_C}} - 1 \right )^{2/3}.
\end{equation}

With this model and the simulated $A(d)$ values in mind, we plot expected displacement fidelities as a function of $Q$ in Figure \ref{fig:outlook} (right).
The vertical dashed lines represent our current device, with a frequency of roughly $\qty{5}{\giga\hertz}$ and a $T_1$ of roughly $\qty{50}{\micro\second}$, and a hypothetical device with an order of magnitude larger $T_1$, representing the current state of the art \cite{wang_millisecond_2022}.
The black dot is our experimentally measured fidelity for $d = 8$ in the absence of pulse corrections.

In this analysis we have considered standard transmon devices only, and have restricted $E_J / E_C$ to a value appropriate for 8-dimensional qudits.
In practice one would ideally choose the device parameters that allow for the desired qudit dimension while also maximizing anharmonicity, which would further reduce coherent errors for $d < 8$.
Furthermore, we have not considered the pulse corrections discussed in the main text, and in that sense our model acts as an upper bound on the achievable error given a particular device.
Finally, we note that alternative device designs such as the inductively shunted transmon \cite{hassani_ist_2023} may be able to encode high-dimensional, long-coherence-time qudits without the restriction on anharmonicity inherent to high-$E_J / E_C$ transmons.
%\bibliography{Bibliography}
%\bibliographystyle{naturemag}

%\end{document}
\end{document}